\documentstyle[aps,preprint,tighten]{revtex}

\begin{document}
\draft
\preprint{BARI - TH 145/93}
\date{June 8, 1993}
\title{
Comment on ``Savvidy ferromagnetic vacuum in three-dimensional
lattice gauge theory'', P.R.L. {\bf 70}, 2053 (1993)}
\author{Paolo Cea$^{a,b}$ and Leonardo Cosmai$^b$}
\address{
$^a$Dipartimento di Fisica dell'Universit\`a di Bari,
70126 Bari, Italy\\
{\rm and}\\
$^b$Istituto Nazionale di Fisica Nucleare, Sezione di Bari,
70126 Bari, Italy\\
(E-mail: cea@bari.infn.it, cosmai@bari.infn.it)
}
\maketitle
\begin{abstract}
We analyze SU(2) gauge theory in a constant chromomagnetic field
in three dimensions. Our analysis instead of supporting the existence
of a non-trivial minimum in the effective potential, corroborates
the evidence of the unstable modes on the lattice.

\vspace{1cm}
\noindent
hep-lat/9306007
\end{abstract}

\vspace{1cm}

\narrowtext

In a recent Letter H. D. Trottier and R. M. Woloshyn investigated
three-dimensional SU(2) lattice gauge theory in an external abelian
chromomagnetic field. The background field is induced by means
of an external current. In the Euclidean continuum the background
action they used
\begin{equation}
\label{eq:1}
S_B = -\frac{1}{2} \int d^3x \, F^{\text{ext}}_{\mu \nu}(x)
\, \left[ \partial_\mu A_\nu^3(x) - \partial_\nu A_\mu^3(x) \right]
\end{equation}
coincides with the one proposed by us in Ref.~\cite{Cea91}
On the lattice
one must discretize the Abelian-like field strength tensor
$F^{\text{A}}_{\mu \nu} = \partial_\mu A_\nu^3(x) - \partial_\nu A_\mu^3(x)$.
Note that the lattice action $S_{\text{B}}$ depends on the discretization
of the Abelian field strength.
The lattice vacuum energy density at zero temperature
in presence of the external magnetic field  $F^{\text{ext}}_{12}$ is given by
\begin{equation}
\label{eq:2}
E \left( F^{\text{ext}}_{12} \right) =
\beta \left[ P_s \left( F^{\text{ext}}_{12}  \right) -
P_t \left( F^{\text{ext}}_{12}  \right) \right]    \; ,
\end{equation}
so that
\begin{equation}
\label{eq:3}
\Delta E \left( F^{\text{ext}}_{12}  \right) =
E \left( F^{\text{ext}}_{12}  \right) - E(0)   \; .
\end{equation}
The authors of Ref.~\cite{Trottier93} used the discretization:
\begin{equation}
\label{eq:5}
F^{\text{A}}_{\mu \nu} = \sqrt{\beta} \, \text{Tr }
\left\{ \frac{\sigma_3}{2i} \left( U_{\mu \nu} -
\left[ U_\mu, U_\nu \right] \right) \right\}  \;,
\end{equation}
and found evidence of a non~trivial minimum for the vacuum energy
difference Eq.(\ref{eq:3}) in the weak field strength region
$x \lesssim 1.5$ ($x=F^{\text{ext}}_{12}/g^3$).

Very recently~\cite{Cea93} we have investigated the same problem on lattices
with periodic boundary conditions. The main differences between our
method and that one of Ref.~\cite{Trottier93} resides in the discretization
of the Abelian field-strength tensor. In Ref.\cite{Cea93}
we define the Abelian magnetic field by means of the Abelian projection.
However we do not found
evidence of the non-trivial minimum. We do not expect that
in the weak external field strength
region the vacuum energy density should display a dramatic dependence
on the discretization of the Abelian field-strength tensor.
Hence we have performed simulations by using the
discretization~(\ref{eq:5}) with the same lattice and statistics as
in Ref.~\cite{Trottier93}. We measured the vacuum energy Eq.~(\ref{eq:3})
and the results are displayed in Fig.~1a. Figure~1a should be compared with
Fig.~3 of Ref.~\cite{Trottier93}.
Unlike Ref.~\cite{Trottier93} we do not found a clear signal for
$x \lesssim 1.5$.
Thus we feel that the non-trivial minimum in Ref.~\cite{Trottier93}
is a lattice artefact. As a further check, we reanalyzed
the data by evaluating directly during Monte Carlo runs the difference
$\left\langle P_s - P_t \right\rangle$. We have checked
that both methods agree for the discretization adopted in Ref.~\cite{Cea93}
and for the U(1) lattice gauge theory.

The results are displayed in Fig.~1b together with the one-loop effective
potential (dotted line), and the vacuum energy density after stabilization
of the unstable modes (dashed line), Eq.(6.25) of Ref.~\cite{Cea93}.

We see that the vacuum energy density difference is positive in
agreement with Ref.~\cite{Cea93}. Moreover the data are quite close to
Eq.~(6.25) of Ref.~\cite{Cea93}. Thus we conclue that the discretization of
Ref.~\cite{Trottier93}, instead of supporting the existence of a non-trivial
minimum in the effective potential, corroborates the evidence
of the unstable modes on the lattice.


%


\begin{figure}
\caption{Vacuum energy density versus $x$.
The vacuum energy difference is evaluated by means of
Eq.~(2) (a),  and directly during Monte Carlo runs (b).
Circles correspond to $\beta=7$, squares to $\beta=10$, and
triangles to $\beta=12$.}
\end{figure}

\end{document}